\documentclass[useAMS,usenatbib]{mn2e}
\usepackage{xspace}
\usepackage[]{natbib}
\usepackage{times}
\usepackage{hyperref} 
\usepackage{amsmath} 
\usepackage{amssymb} 
\usepackage{graphicx} 
\usepackage{indentfirst} 
\usepackage{pdflscape} 
\usepackage{placeins} 
\usepackage{nicefrac} 
\usepackage{afterpage} 
\usepackage{rotating} 
\usepackage{booktabs} 
\usepackage{rotating} 
\usepackage{color} 
\usepackage{caption} 
\usepackage{pdfpages}

\usepackage{fancyvrb} 

 \newcommand{\bvp}{\texttt{BayesVP\,}}
 
\begin{document}

\title[Bayesian Voigt Profile Fitting]{\texttt{BayesVP}: a Bayesian Voigt profile fitting package}

\author[Liang \& Kravtsov]{Cameron J. Liang$^{1}$\thanks{E-mail:jwliang@oddjob.uchicago.edu} and Andrey V. Kravtsov$^{1,2}$\\
\\
$^{1}$Department of Astronomy \& Astrophysics, and Kavli Institute for Cosmological Physics, University of Chicago, Chicago IL 60637 \\
$^{2}$Enrico Fermi Institute, The University of Chicago, Chicago, IL 60637, USA}

\pagerange{\pageref{firstpage}--\pageref{lastpage}} \pubyear{2015}

\maketitle

\label{firstpage}

\begin{abstract}

We introduce a Bayesian approach for modeling Voigt profiles in absorption spectroscopy and its implementation in the python package, \bvp, publicly available at \href{https://github.com/cameronliang/BayesVP}{\texttt{https://github.com/cameronliang/BayesVP}}. The code fits the absorption line profiles within specified wavelength ranges and generates posterior distributions for the column density, Doppler parameter, and redshifts of the corresponding absorbers. The code uses publicly available efficient parallel sampling packages to sample posterior and thus can be run on parallel platforms.  \bvp supports simultaneous fitting for multiple absorption components in high-dimensional parameter space. We provide other useful utilities in the package, such as explicit specification of priors of model parameters, continuum model, Bayesian model comparison criteria, and posterior sampling convergence check. 

\end{abstract}

\begin{keywords}
 methods:data analysis -- quasars:absorption lines
\end{keywords}

\section{Introduction}
Absorption line spectroscopy is undoubtedly one of the most important tools for probing the physical properties of absorbing gas in a variety of astrophysical environments.  Spectral analyses are ubiquitous in astrophysics and have a long and rich history in studies of stars, galaxies, gas, cosmology and more. 

One of the basic tasks in such analyses is identification and characterization of spectral absorption lines. The most common way to do the latter is to fit the Voigt profile (VP) to the absorption profile and then use its parameters to extract basic properties of the absorbing gas, such as column density, Doppler parameter, and central velocity/redshift. 

A number of codes for modeling absorption lines with the Voigt profile is available publicly. For example, \texttt{VPFIT} is a popular $\chi^2$-based code commonly used in the community over many years \citep{Carswell1991}.  \texttt{autoVP} is an another $\chi^2$ grid-search based code, an automated Voigt profile fitting procedure that chooses the best number of absorption components, designed for Ly$\alpha$ forest statistics \citep{Dave1997}. 

While these codes have provided immense values to the community, there are fundamental limitations in methods based on $\chi^2$ grid-search in the Voigt profile fitting. For example, such methods cannot constrain model parameters of non-detections in a statistically rigorous manner. The methods make very specific assumptions about the likelihood and the nature of the flux errors. In addition, it is challenging to capture the degeneracy between column density and Doppler parameter when absorption lines are saturated.  Therefore, uncertainties of the best-fit parameters can be under- or overestimate estimated in a coarse grid of the parameter space. Furthermore, the grid search becomes prohibitively computationally expensive or impossible when fitting multiple absorption components because the number of parameters and thus the number of dimensions of parameter space is large.  Thus, one cannot reliably and rigorously find the best fit model for physical properties of the gas using the optimal number of absorption components. This can affect our physical interpretation of the kinematics of the gas (i.e., number of velocity components and the total dispersion) and the column density distribution function \citep[for example, see Appendix in][]{Gurvich2017}. 

Motivated by these problems, we developed a Bayesian approach to the Voigt profile fitting combined with efficient algorithms for sampling posterior distribution of parameter values. We introduce an implementation of such method in the python package \texttt{BayesVP},  which simultaneously addresses these issues and provides additional benefits. These include explicit control of the priors of parameters and providing constraints in a form of posterior distributions, which allows uniform treatment of detections and non-detections.  Using efficient parallel sampling methods, \texttt{BayesVP} can fully explore high-dimensional parameter space efficiently, providing more accurate parameters uncertainties and their degeneracies. This allows us to reliably find the best number of absorption components and to include an additional continuum model. One can also explicitly compare models using Bayesian model comparison criteria.  In section 2, we describe the Bayesian approach to absorption line fitting and our specific implementation of this approach in the \bvp package.  We include a short tutorial on the basic usage of the package in the Appendix.  

\section{The Package: \texttt{BayesVP} }

\subsection{Posterior Distribution and sampling}

In the Bayesian approach we  constrain the posterior distribution, $\pi(\vec{\theta})$, of model parameters vector of column density, Doppler broadening factor, and redshift, $ \vec{\theta} = \{N,b,z\}$, given the vector of spectral fluxes and their uncertainties, $\vec{F}$. The joint posterior distribution is simply proportional to the product of likelihood $\mathcal{L}(\vec{F} | \vec{\theta} )$ and the prior pdf of the parameters: 
\begin{equation} 
p(\vec{\theta} | \vec{F}) \propto \mathcal{L}(\vec{\theta} | \vec{F}) p(\vec{\theta}), 
\end{equation}
where the total likelihood $\mathcal{L}$ is the product of all likelihoods of the individual pixels within a spectral segment of interests: 
\begin{equation} \mathcal{L}(\vec{\theta} | \vec{F}) = \prod_{i} \ell_i(\vec{\theta} | F_i). \end{equation}
Using methods, such as Markov Chain Monte Carlo (MCMC), that allow sampling of unnormalized distributions, we can
sample the unnormalized posterior: 
\begin{equation} 
\pi(\vec{\theta} | \vec{F}) = \mathcal{L}(\vec{\theta} | \vec{F}) p(\vec{\theta}), 
\label{eq:posterior}
\end{equation}

The individual likelihood $l_i$ of a given pixel $i$ at $F_i$ depends on the noise model. The specific noise characteristic depends on properties of telescopes and their instruments. In the limit of a large number of photons, we usually assume that $l_i$ is a Gaussian: 
\begin{equation} \ell_i ( \vec{\theta} | F_i) = \frac{1}{\sqrt{2 \pi \sigma_i^2}} \exp \left[-\frac{(\hat{F_i}(\vec{\theta}) - F_i)^2}{2 \sigma_i^2} \right]\end{equation}
where $\sigma_i$ is the uncertainty of the flux $F_i$ at pixel $i$.  However, the noise model can be modified and a different likelihood can be used for low photon counts (e.g., Poisson distribution). 
Although currently \bvp implements only Gaussian likelihood, this can be easily changed to a different likelihood by supplying a different input log-likelihood function to the {\tt Posterior} class in {\tt Likelihood.py}. In \bvp, we also provide an easy way to specify explicit priors based on knowledge of the specific problem in mind.  

We sample the posterior distribution defined by eq. \ref{eq:posterior} using an efficient sampling method. This is particularly useful in the Voigt profile fitting because it is common to fit multiple absorption components simultaneously. The number of parameters can thus grow quickly, since each absorption component adds three parameters (e.g., $N$, $b$, $z$). In addition, many sampling methods can be easily parallelized, which allows  efficient sampling of the high-dimensional parameter space.  Due to blending between components and the relationship between Doppler parameters and column densities in the saturated regime, Voigt profile parameters can be highly correlated.  The simplest Metropolis-Hasting MCMC algorithm that samples the parameter space with isotropic steps is thus not optimal. We thus use two MCMC sampling algorithms designed to efficiently sample narrow distributions  of highly correlated parameters and in high dimensional spaces. In particular, \bvp is set up to use the affine-invariant ensemble sampling algorithm of \citep{GoodmanWeare2010}  implemented in the \texttt{emcee} python package \citep{ForemanMackey2013} \footnote{\url{https://github.com/dfm/emcee}}, and an additional powerful kernel-density-estimate based MCMC sampler \texttt{kombine} \footnote{\url{https://github.com/bfarr/kombine}} \citep{kombine}. 

\subsection{Voigt Profile Model}

The Voigt profile is parameterized by the absorber column density $N$, Doppler parameter $b$, and redshift $z$. The normalized flux at frequency $\nu$ (or wavelength $\lambda$) is simply, 
\begin{equation} F(\nu | N, b, z) =  \exp[-\tau(\nu)]  \end{equation}
where the optical depth $\tau$ at frequency $\nu$ of a system with $(N, b, z)$ is: 
\begin{equation} \tau(\nu | N, b, z)  = N \sigma_0 f_{\rm osc} \Phi(\nu | b, z)  \end{equation}
and $\Phi(\nu)$ is the convolution of a Gaussian profile (due to the Doppler broadening) and a Lorentzian profile (due to pressure broadening). For computation speed, we compute the Voigt profile function $\Phi(\nu)$ via the real part of the Faddeeva function $w(x + i y)$ routine in the \texttt{SciPy} package\footnote{\url{https://docs.scipy.org/doc/scipy/reference/generated/scipy.special.wofz.html}\\ 
\url{http://ab-initio.mit.edu/wiki/index.php/Faddeeva_Package}}:
\begin{equation} \Phi(\nu | b, z) = \frac{\Re[w(x+i y)]}{\sqrt{\pi}\nu_D}   \end{equation}
where $x = \Delta \nu/ \Delta \nu_D$ and $y = \Gamma / (4 \pi \Delta \nu_D)$. Also, $\Delta \nu = \nu - \nu_0 (1+z)^{-1}$, $\Delta \nu_D = b \nu / c$. In addition, $\sigma_0 = \frac{\sqrt{\pi} e^{2}}{m_e c^2}$ is the cross section. 

Collectively, the set of atomic constants $\{e, m_e, f_{\rm osc}, \nu_0, \Gamma\}$ are the charge and mass of the electron, oscillator strength, rest-frame frequency and damping coefficient of an atomic transition. We include atomic parameters from the UV to optical wavelength regimes compiled from \cite{Morton2003}. Any additional transitions can be easily added in the supplementary data file included in \bvp.

\subsection{Continuum model}
We implement a polynomial continuum model in addition to the Voigt profile model fitting. The addition of the continuum model moves some of the error budget from systematic uncertainties to statistical uncertainties. The continuum is parametrized as a polynomial in \texttt{BayesVP}: 
\begin{equation} C(\lambda | a_i) = \sum_{i}  a_i (\lambda -  \bar{\lambda} )^i  + \bar{F} \end{equation}
\begin{equation} F_{\rm model}(\lambda) = C (\lambda | a_i) \times F_{\rm Voigt} (\lambda | N, b, z) \end{equation}
where $\{a_i\}$ are the continuum parameters.  Although \bvp can fit a high order polynomial,  in our experience for typical wavelength ranges used in absorption line fits, linear or quadratic polynomial 
are sufficient to model continuum accurately. Thus, we recommend fitting up to a quadratic polynomial (i.e., three continuum parameters). In any case, the degree of the polynomial should be considerably smaller than the number of spectrum pixels dominated by continuum.  

\begin{figure}
\begin{center}
\includegraphics[scale=0.33]{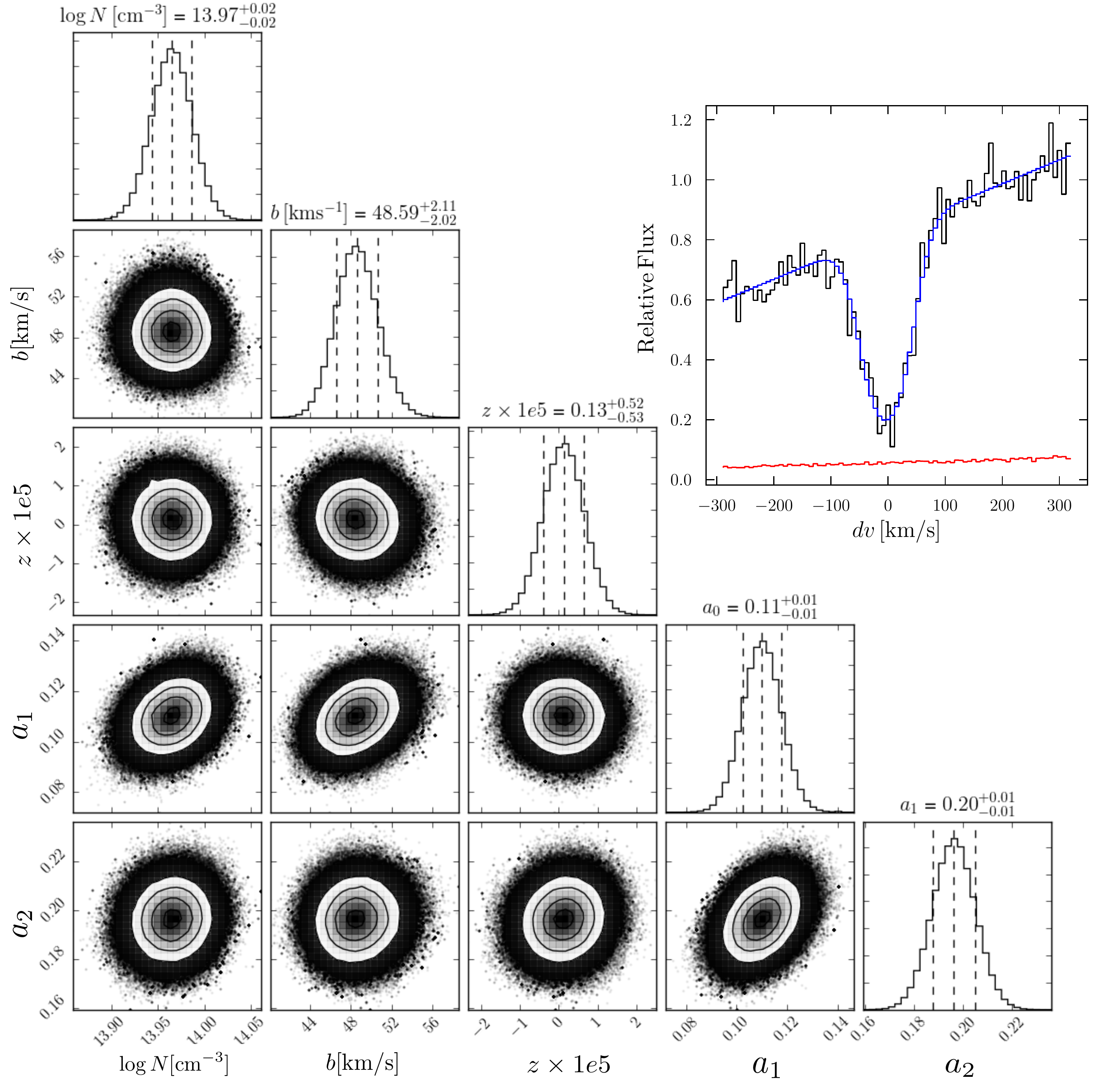}
\caption{Marginalized posterior distributions for parameters in a single component Voigt profile fit with a linear continuum model fit to a synthetic spectrum. The top right panel shows the best fit model relative to the input spectrum over the fitted wavelength interval.  \label{fig:cont_model}}
\end{center}
\end{figure}

Note that we have pivoted the polynomial fit from the observed wavelength $\lambda$ to $(\lambda -  \bar{\lambda} )$, where $\bar{\lambda}$ and $\bar{F}$ are the median wavelength and flux of the input data. In Figure \ref{fig:cont_model}, we show a test that includes a continuum model in a synthetic absorption line, where \bvp successfully captures the correct input parameters.

\subsection{Convergence Criteria}

\begin{figure}
\begin{center}
\includegraphics[scale=0.75]{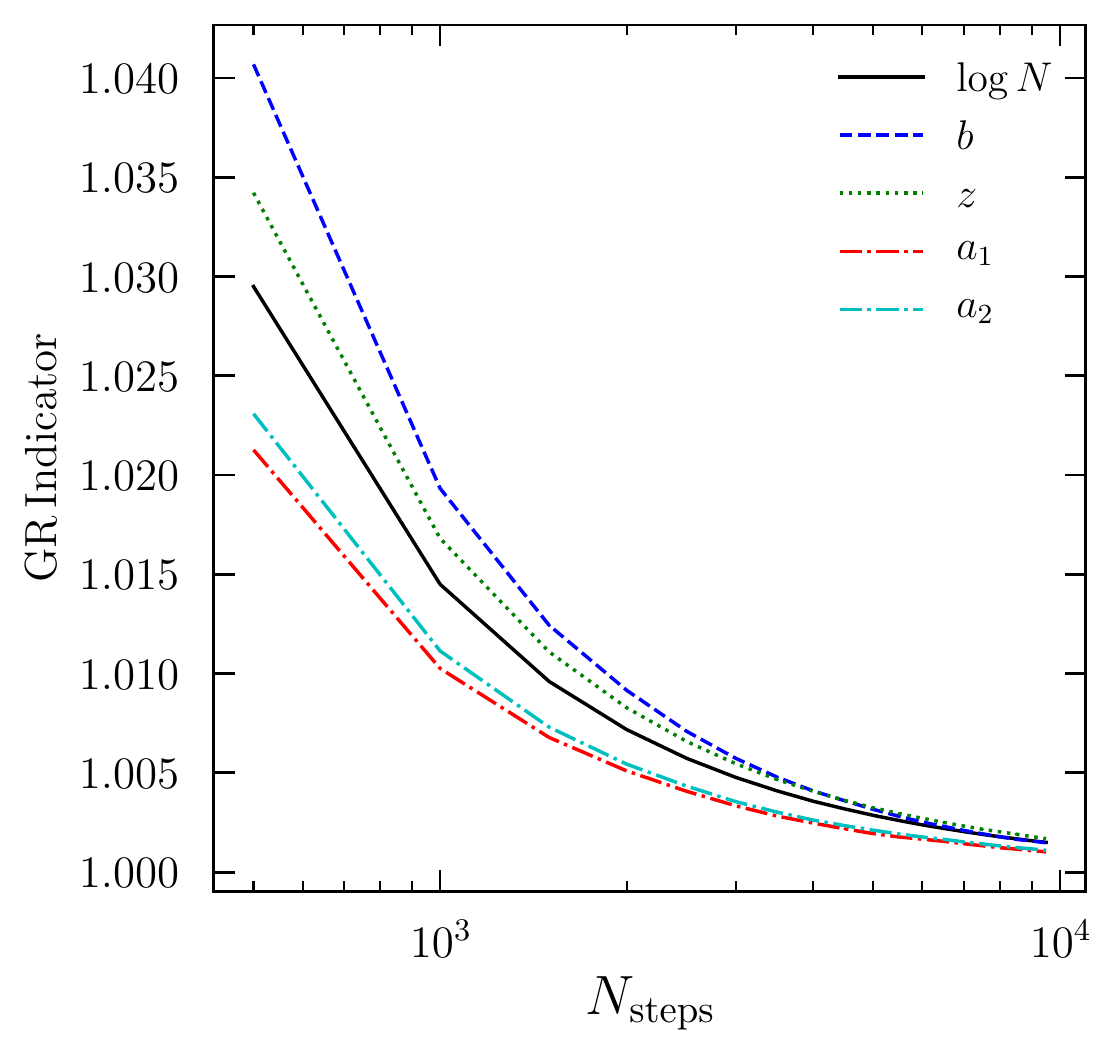}
\caption{Evolution of the Gelman-Rubin (GR) indicator for model parameters as a function of MCMC steps. A perfect convergence of a parameter corresponds a value of one. Users may choose a level of convergence based on a threshold of GR indicator to decide the number of MCMC steps necessary to run. Note also that parameters converge at different rates. \label{fig:gr_ind}}
\end{center}
\end{figure}

When sampling a posterior, it is important to demonstrate the convergence of sample chains in order to trust the resulting posterior and statistics derived from it. \texttt{BayesVP} package provides functionality for convergence tests using the Gelman-Rubin indicator. The basic idea is that the chains should reach a state where the parameter distributions are stationary. To do so, the GR indicator compares the variances of parameters within and across independently sampled chains. We adopt the definition used in \cite{GelmanRubin1992}: 
\begin{equation}
R = \frac{V}{W} = \frac{N_w + 1}{N_w} \frac{\sigma^2_+}{W} - \frac{N_s - 1}{N_w N_s}
\end{equation}
where $W$ and $V$ are the ``within-chain" variance and ``between-chain" variance. And  $\sigma^2_+$ is the estimator for the true variance of a parameter. In addition, $N_w$ and $N_s$ are the number of independent chains (or ``walkers'')  and the number of steps (length of the individual chains), respectively.  A perfect convergence would correspond to the ``within-chain" and ``between-chain" variances matching each other ($R = 1$). Figure \ref{fig:gr_ind} shows an example the GR indicator as a function of the steps for the parameters in the continuum model fit shown in Figure \ref{fig:cont_model}.

\subsection{Model Comparison}
\begin{figure}
\begin{center}
\includegraphics[scale=0.33]{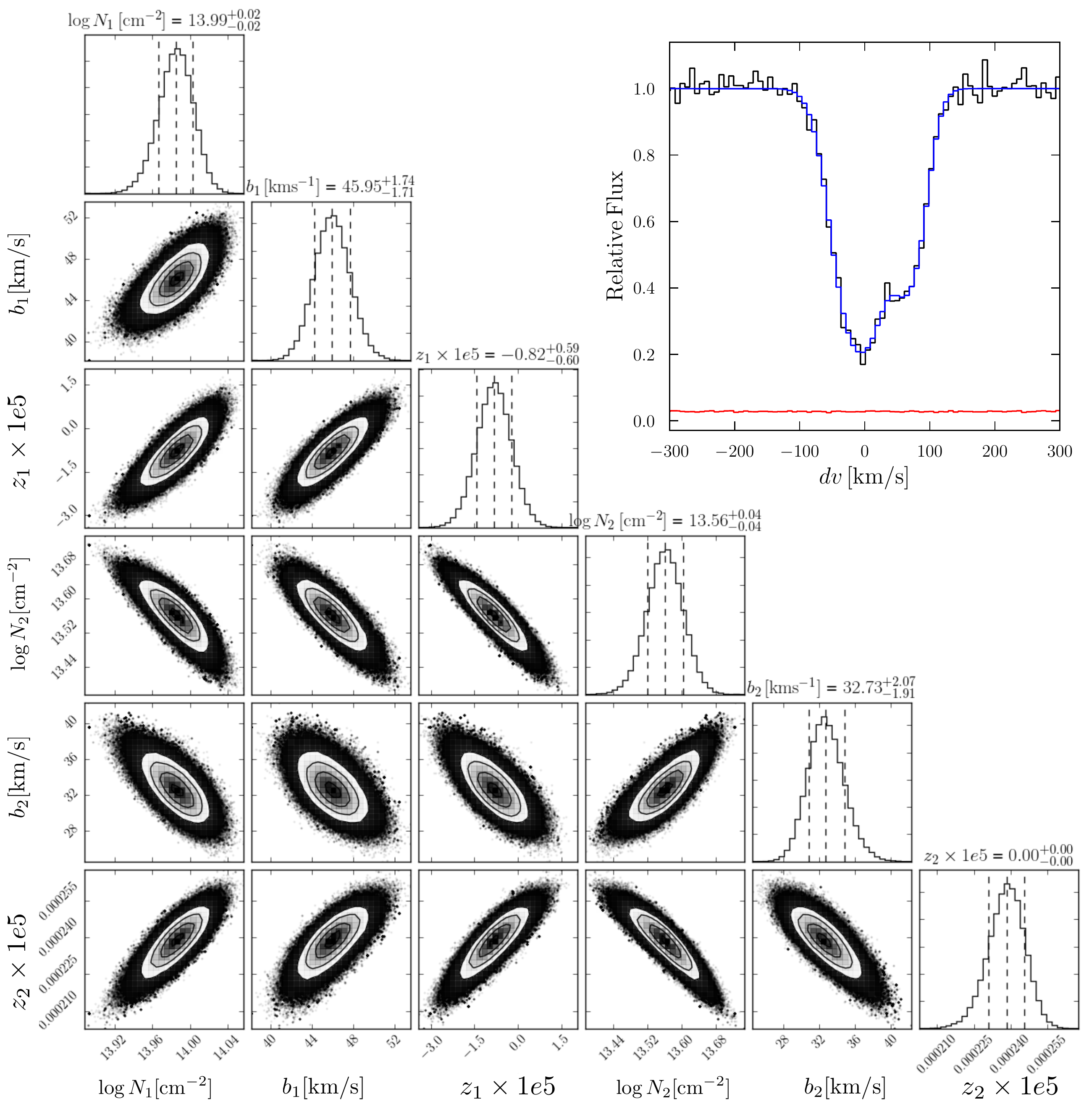}
\caption{This figure shows the posterior distributions Voigt profile fit of a synthetic spectrum. This shows that a 2-component model is a good fit to the data. Note also the parameters of the two Voigt profiles are correlated. \label{fig:multi_comp}}
\end{center}
\end{figure}

\begin{figure}
\begin{center}
\includegraphics[scale=0.75]{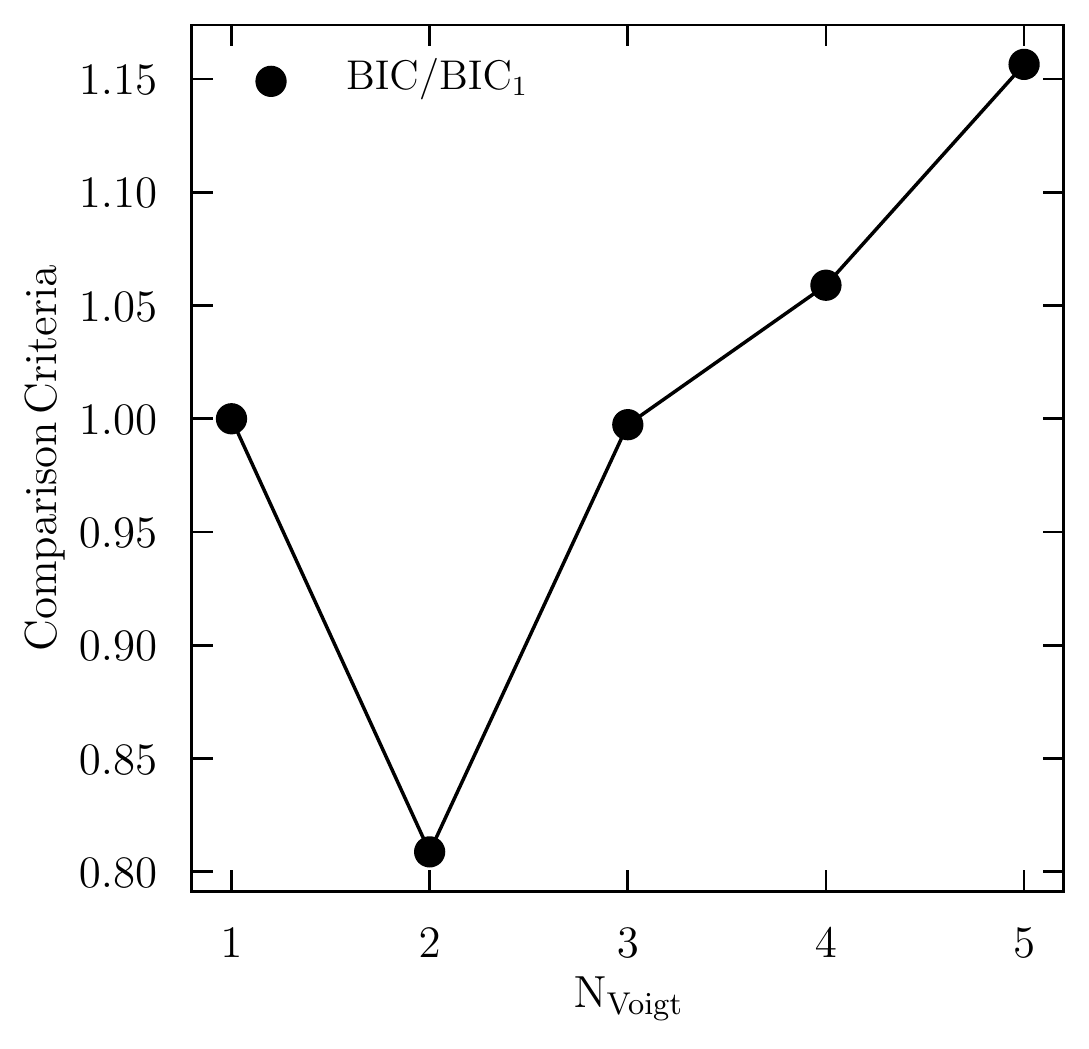}
\caption{Model comparison based on $BIC$ as a function of the number of Voigt components ($N_{\rm Voigt}$) for the spectrum shown in Figure \ref{fig:multi_comp}. Note that there are three free parameters in each component. The model of 2 components with the minimum value is preferred. \label{fig:bic}}
\end{center}
\end{figure}

Observed absorption lines are frequently blended. The best number of absorption components for the observed profiles is not always clear.  The column densities and Doppler parameters are strongly affected by the number of components that we choose to fit the data. Our physical interpretation of the absorbers may also change depending on the number of velocity components. One of the main advantages of a Bayesian approach is the ability to select the best fit model based on some objective criteria when it is not clear how many components one should choose. In \bvp, we implement some measures for choosing the best fit models:  the Bayesian information criterion ($BIC$) for a given model $M$ \citep{Ivezi2014}:
\begin{equation} BIC = - 2 \ln (L^0 (M)) + 2k + k \ln N \end{equation}
where, $L^0(M)$ denotes the maximum value of the likelihood of model $M$, and $k$ is the number of model parameters and $N$ is the number of points in the MCMC chains. The objective for the best-fit model is to minimize $BIC$.

In general, odds ratio is also useful for comparison of two models $M_1$ and $M_2$: 
\begin{equation} O_{21} =  L(M_2)  / L(M_1)\end{equation}
For this measure, model 2 is preferred if $O_{21} \gg 1$. However, it is computationally expensive to compute $L(M)$ because it involves integrating the posterior distribution over all $k$ model parameters, $L(M) = \int \pi (\vec{\theta}) d^k \theta$. Instead, we calculate $L(M)$ using the local density estimate of the points sampled by MCMC chains $\rho(\vec{\theta})$ and the posterior $\pi (\vec{\theta})$, i.e. $L(M) = N \pi(\vec{\theta})/\rho(\vec{\theta})$ \citep[see section 5 in ][for more details.]{Ivezi2014}

The basic idea for these criteria (e.g., $BIC$) is that the models with a larger number of parameters are penalized and have to compensate with the larger values of $L(M)$ or $L^0(M)$.  As an example, in Figure 3  we show results of a two-component Voigt fit for a blended absorption profile (with two input absorption components) and the corresponding posterior distributions of the model parameters. Clearly, the two-component fit describes the observed flux quite well. The choice of a two-component fit is also supported by the comparison criteria shown in Figure \ref{fig:bic}.

\subsection{Uniform treatment for upper limits, lower limits and detections}

Absorption line fits for weak lines in noisy spectra often result in unconstrained parameters. In the $\chi^2$ based analyses, these are usually treated as upper or lower limits.  For example, in observational analyses of the absorption profiles of the circumgalactic medium, lines of sights of large impact parameters usually result only in upper limits \citep[e.g.,][]{LiangChen2014}. The inclusion of such limits when fitting a model of the column density profile or comparing to simulation results \citep[e.g.,][]{Liang2016} is challenging and requires strong assumptions about their posterior form. 

Thus, there has been a substantial effort in trying to model the probability of the column density of upper-limit systems in order to maximize their utility, such as constraining the gas properties \citep[e.g., density and metallicity; ][]{Crighton2015, Fumagalli2016, Stern2016}. On the other hand, when the absorption lines are saturated, it is difficult to constrain the column density precisely due to its degeneracy with the Doppler parameter. In this case, lower limits are often quoted in the literature. An improvement over the lower limits is bracketed ranges using the absence of damping wings \citep[e.g.,][]{Johnson2015}. In both of these cases, when the original data is available, this probability distribution can be directly measured as a marginalized posterior. In this case, no assumptions about the shape of the probability distribution are needed. We refer readers to Figure 6 in our companion paper \citep{Liang2017} to see the diversity of the posterior distributions for the cases that would usually be identified as  
upper or lower limits and compared to the posterior distributions of ``detections.''

\section{Summary}

We present a python package \bvp, which implements a Bayesian approach for modeling absorption lines with the Voigt profile. The package can be used to simultaneously fit multiple absorption
components to constrain column density, Doppler parameter, and redshifts of the absorbing gas. The parameter constraints are derived in the form of marginalized posterior distributions, which allows for uniform treatment of the non-detections (i.e., upper  limits) and detections in subsequent analyses, model fits, and comparisons of observations with simulations. The \bvp package is set up to use the affine-invariant MCMC method implemented in the MCMC {\tt emcee} package and the kernel-density-estimate-based ensemble sampler {\tt KOMBINE}  to efficiently sample the posterior distributions of highly correlated parameters. It also provides other useful utilities, such as convolution with instrumental line spread function (LSF), explicit control of parameter priors, Bayesian model comparison criteria, and sampling convergence check. The \bvp package is designed to have a simple user interface with a single configuration file that users can explicitly define the line spread function, the number of walkers, MCMC steps, and parallel threads. \bvp is publicly available at \texttt{https://github.com/cameronliang/BayesVP}. In the Appendix below we present an example showing how to use the package to fit absorption lines.

\section*{Acknowledgments}

CL and AK were supported by a NASA ATP grant NNH12ZDA001N, NSF grant AST-1412107, and by the Kavli Institute for Cosmological Physics at the University of Chicago through grant PHY-1125897 and an endowment from the Kavli Foundation and its founder Fred Kavli. CL is partially supported by NASA Headquarters under the NASA Earth and Space Science Fellowship Program - Grant NNX15AR86H.



\section*{Appendix}
We describe the basic usage of the package in this section. \texttt{BayesVP} is meant to run with a configuration file in background, as it can take a few minutes for MCMC sampling, depending on the chosen number of walkers, steps, and parallel processes. This section illustrates the basic interactive use of the code and setting up a configuration file. 

For this example, we will explicitly change the path to the location of the package for importing \bvp. 

\begin{verbatim}
In [1]: 
import sys
sys.path.append(`/Users/cameronliang/BayesVP')
\end{verbatim}
We can now import \texttt{BayesVP}. Let us also import an object, \texttt{WriteConfig}, that interactively asks the user a few questions to create a config file.

\begin{verbatim}
In [2]: import BayesVP, WriteConfig
\end{verbatim}

Let us assume that the spectrum in question is located in the following directory:

\begin{verbatim}
In [3]: spectrum_path = `/Users/cameronliang/
BVP_tutorial/spectrum'
\end{verbatim}

The file name of the example spectrum is OVI.spec, with three of columns of data: wave, flux, error. We can use the \texttt{WriteConfig} routine to set up the config file like so:

\begin{verbatim}
In [4]: config = WriteConfig.interactive_QnA()
        config.QnA()

Path to spectrum:
/Users/cameronliang/BVP_tutorial/spectrum/

Spectrum filename: OVI.spec
filename for output chain: o6
atom: O
state: VI
Maximum number of components to try: 1
Starting wavelength(A): 1030
Ending wavelength(A): 1033

Enter the priors:
min logN [cm^-2] = 10
max logN [cm^-2] = 18
min b [km/s] = 0
max b [km/s] = 100
central redshift = 0
velocity range [km/s] = 300

Enter the MCMC parameters:
Number of walkers: 400
Number of steps:  2000
Number of processes: 8
Model selection method bic(default),aic,bf: bic
MCMC sampler kombine(default), emcee: kombine
\end{verbatim}

\begin{verbatim}
Written config file: /Users/cameronliang/
BVP_tutorial/spectrum/bvp_configs/config_OVI.dat
\end{verbatim}

The config file is automatically written within a subdirectory where the spectrum is located. We can now run the MCMC fit shown below. Note that \bvp can be run by supplying the full path to the config file as a command line argument.  \bvp will print to screen the relevant information from the config file. In this example, we are fitting an O\,VI transition with rest wavelength of 1031.926 \AA.

\begin{verbatim}
In [5]: config_fname = spectrum_path + 
	'/bvp_configs/config_OVI.dat'
	
In [6]: BayesVP.bvp_mcmc(config_fname)
	
--------------------------------------------
Config file: /Users/cameronliang/BVP_tutorial/
           spectrum/bvp_configs/config_OVI.dat
--------------------------------------------

Spectrum Path:/Users/cameronliang/BVP_tutorial/
           spectrum/
Spectrum name: OVI.spec
Fitting 1 component(s) with transitions:
    Transition Wavelength: 1031.926
Selected data wavelegnth region:
    [1030.000, 1033.000]
MCMC Sampler: kombine
Model selection method: bic
Walkers,steps,threads : 400,2000,8
Priors:
logN:     [min, max] = [10.000, 18.000]
b:        [min, max] = [0.000, 100.000]
redshift: [min, max] = [-0.00100, 0.00100]

Written chain: /Users/cameronliang/BVP_tutorial/
          spectrum/bvp_chains_0.000000/o6.npy
--------------------------------------------
\end{verbatim}

We complete the fitting process after the step above since the output chain (a binary file ends with .npy) contains all the information that we need. We would like to plot the results and 
write the best fit spectrum into a file. There are tools in the package that can help us do so:

\begin{Verbatim}
In [7]: import PlotModel as pm
	from Config import DefineParams
\end{Verbatim}

We first extract all of the relevant information using the DefineParams function:
\begin{Verbatim}
In [8]: config_params = DefineParams(config_fname)
--------------------------------------------
Config file: /Users/cameronliang/BVP_tutorial/
	spectrum/bvp_configs/config_OVI.dat
\end{Verbatim}

\begin{Verbatim}
In [9]: output = pm.ProcessModel(config_params)
In [10]: redshift = 0.0; dv = 300;
         output.plot_model_comparison(redshift,dv)
         output.write_model_summary()
         output.write_model_spectrum()
         output.plot_gr_indicator()
         output.corner_plot()
\end{Verbatim}

\begin{figure}
\begin{center}
\includegraphics[scale=0.45]{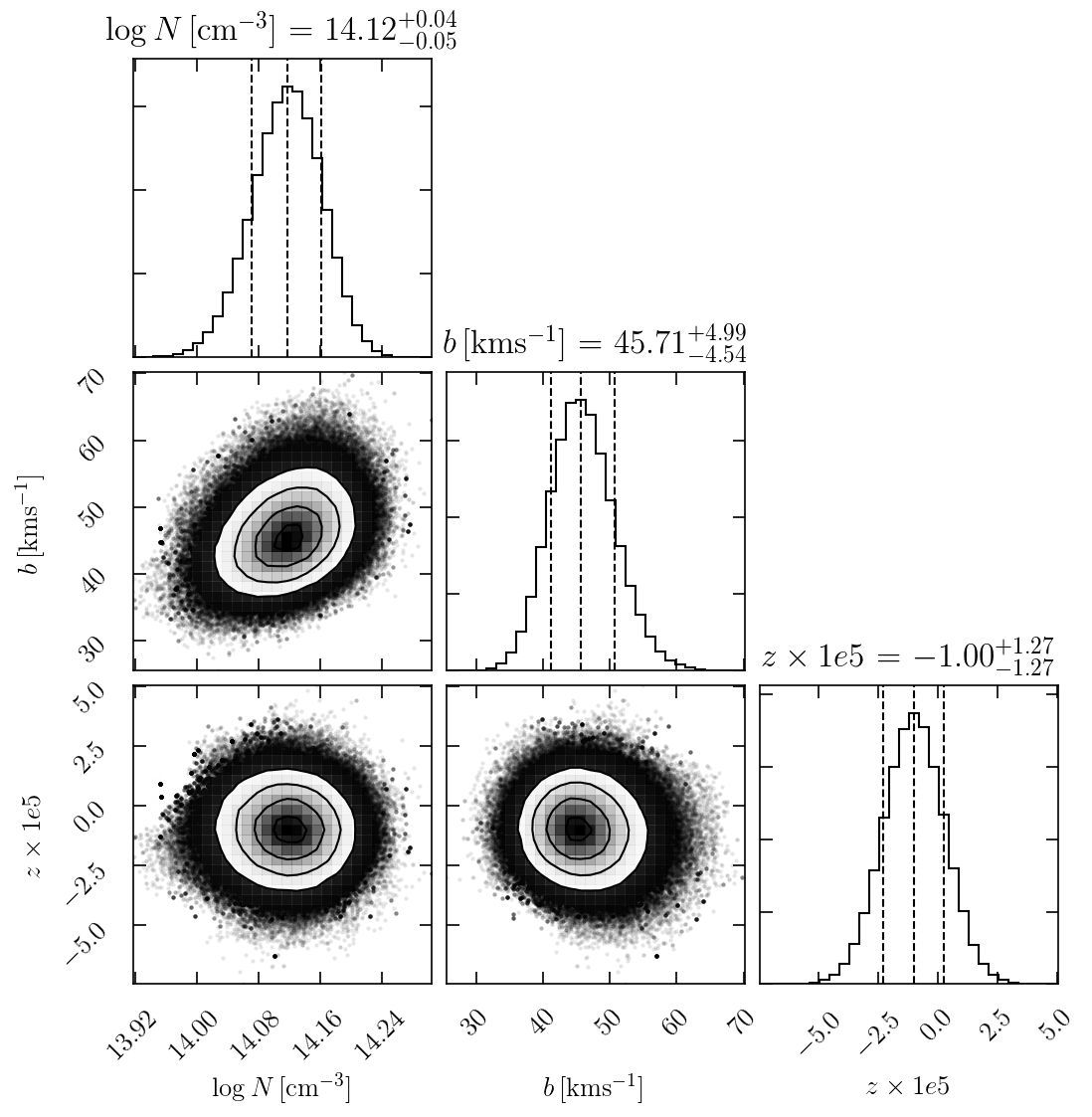}
\caption{Posterior distribution for the example fit. \label{fig:ex_o6_corner}}
\end{center}
\end{figure}

\begin{figure}
\begin{center}
\includegraphics[scale=0.7]{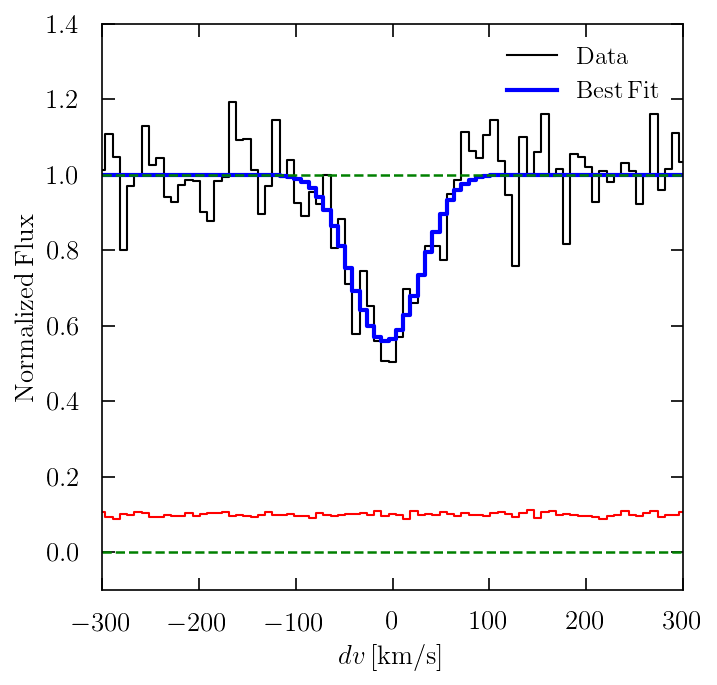}
\caption{Best fit (blue) and original data (black) for the example fit. \label{fig:ex_o6_fit}}
\end{center}
\end{figure}

The data products will be written in a subdirectory \texttt{processed\_products\_0.000000}, where the redshift of the system is appended to the subdirectory name. The final posterior and the best fit data are shown in Figure \ref{fig:ex_o6_corner} and Figure \ref{fig:ex_o6_fit}. 

\bibliographystyle{mn}
\bibliography{ms}

\end{document}